\documentclass[final,5p,times,twocolumn]{elsarticle}

\usepackage{graphicx}

\usepackage{amssymb}

\usepackage{subfigure}

\journal{Physics Letters B}

\begin{document}

\begin{frontmatter}



\title{New Precision Measurement of Hyperfine Splitting of Positronium}


\author[tokyo]{A.~Ishida\corref{cor1}}
\ead{ishida@icepp.s.u-tokyo.ac.jp}

\author[tokyo]{T.~Namba}
\author[tokyo]{S.~Asai}
\author[tokyo]{T.~Kobayashi}
\author[komaba]{H.~Saito}
\author[kek]{M.~Yoshida}
\author[kek]{K.~Tanaka}
\author[kek]{and A.~Yamamoto}

\address[tokyo]{Department of Physics, Graduate School of Science, and International Center for Elementary Particle Physics (ICEPP), The University of Tokyo, 7-3-1 Hongo, Bunkyo-ku, Tokyo 113-0033, Japan}
\address[komaba]{Department of General Systems Studies, Graduate School of Arts and Sciences, The University of Tokyo, 3-8-1 Komaba, Meguro-ku, Tokyo 153-8902, Japan}
\address[kek]{High Energy Accelerator Research Organization (KEK), 1-1 Oho, Tsukuba, Ibaraki 305-0801, Japan}

\cortext[cor1]{Corresponding author (TEL:+81-3-3815-8384 / FAX:+81-3-3814-8806)}

\begin{abstract}
The ground state hyperfine splitting of positronium $\Delta _{\mathrm{HFS}}$
is sensitive to high order corrections of quantum electrodynamics (QED) in bound state.
The theoretical prediction and the averaged experimental value for 
$\Delta _{\mathrm{HFS}}$ has a discrepancy of 15 ppm, which is equivalent to 3.9 standard deviations (s.d.).
A new precision measurement which reduces the systematic uncertainty from the positronium thermalization effect was 
performed, in which the non-thermalization effect was measured to be as large as $10 \pm 2\,{\mathrm{ppm}}$ in a timing window 
we used. 
When this effect is taken into account, our new result becomes 
$\Delta _{\mathrm{HFS}} = 203.394\,2 \pm 0.001\,6 ({\mathrm{stat., 8.0\,ppm}}) \pm 0.001\,3 ({\mathrm{sys., 6.4\,ppm}})$\,GHz, 
which favors the QED prediction within 1.2 s.d. and disfavors the previous experimental average by 2.6 s.d.
\end{abstract}


\end{frontmatter}


\section{Introduction}
\label{sec:introduction}
Positronium (Ps), a bound state of an electron and a positron, 
is a purely leptonic system 
which allows for sensitive tests of quantum electrodynamics (QED) in bound state.
Orthopositronium (o-Ps, 1$^3S_1$) decays mostly into three $\gamma$ rays with a decay rate of 
$\Gamma_{{\mathrm{o}}{\textrm{-}}{\mathrm{Ps}}} = 7.040\,1(7)\,\mu\mathrm{s}^{-1}$~\cite{KATAOKA}.
On the other hand, parapositronium (p-Ps, 1$^1S_0$) decays mostly into two $\gamma$ rays 
with a decay rate of $\Gamma _{{\mathrm{p}}{\textrm{-}}{\mathrm{Ps}}} = 7.990\,9(17)\,\mathrm{ns}^{-1}$~\cite{GAMMA-0}.  
The ground state hyperfine splitting between 
o-Ps and p-Ps (Ps-HFS, $\Delta _{\mathrm{HFS}}$) is an ideal probe for the precise test of the 
bound-state QED.
The combined value of the two 
most precise experiments is 
$\Delta _{\mathrm{HFS}} ^{\mathrm{exp}} = 
203.388\,65(67)\,\mathrm{GHz} \,(3.3\,\mathrm{ppm})$
~\cite{MILLS-I,MILLS-II,HUGHES-V}. 
Recent developments in non-relativistic QED (NRQED) have added 
$O(\alpha ^3 \ln \alpha ^{-1})$ corrections to the 
theoretical prediction which now stands at 
$\Delta_{\mathrm{HFS}} ^{\mathrm{th}} = 
203.391\,69(41)\,\mathrm{GHz}\,(2.0\,\mathrm{ppm})$
~\cite{HFS-ORDER3-KNIEHL,HFS-ORDER3-MELNIKOV,HFS-ORDER3-HILL}. 
A discrepancy of 3.04(79)\,MHz (15\,ppm), which is equivalent to 3.9 standard deviations (s.d.), 
between $\Delta _{\mathrm{HFS}} ^{\mathrm{exp}}$ 
and $\Delta_{\mathrm{HFS}} ^{\mathrm{th}}$ might 
be due to common systematic uncertainties in the previous experiments. 
There are two possible common systematic uncertainties in the previous experiments. 
One is the unthermalized o-Ps contribution\footnote{Ps thermalization is a process that 
Ps loses its kinetic energy from initial energy $E_{0}$ to room temperature.}
which results in underestimation of a material effect. 
This effect has already been shown to be 
significant in the so-called o-Ps lifetime puzzle~\cite{ASAI,JINNAI,OPSMICHIGAN}, 
which was a history of a disagreement of the o-Ps lifetime between experimental values 
and theoretical calculations finally solved by taking into account the effect.
Another source of systematic uncertainties is the possible non-uniformity of the magnetic field which was mentioned as
the most significant systematic uncertainty in the previous experiments. 

All the previous precision measurements were indirectly performed by stimulating the transition of 
the Zeeman splitting ($\Delta _{\mathrm{Zeeman}}$) under a static magnetic field. 
One experiment is trying to measure $\Delta _{\mathrm{HFS}}$ directly, 
but it has not obtained the result yet~\cite{PhysRevLett.108.253401,DIRECT_2}. 
Other independent 
experiments~\cite{Sasaki2011121,PhysRevLett.109.073401}
have not yet reached a sufficient level of precision to address the discrepancy. 
The relationship between $\Delta _{\mathrm{HFS}}$ and $\Delta _{\mathrm{Zeeman}}$ under 
a static magnetic field $B$ is 
approximately given by the Breit-Rabi equation 
\begin{equation}
\Delta _{\mathrm{Zeeman}} \approx \frac{1}{2} \Delta _{\mathrm{HFS}} 
\left( \sqrt{1+4q^2} - 1 \right) ,
\label{eq:Zeeman}
\end{equation}
where $q$ is given as $g^{\prime}\mu _B B  / \left( h\Delta _{\mathrm{HFS}} \right)$, 
$g^{\prime} = g\left( 1-\frac{5}{24} \alpha ^2 \right)$ 
is the $g$ factor of the positron (electron) in Ps~\cite{G-FACTOR-1,G-FACTOR-2,G-FACTOR-3,G-FACTOR-4}, 
$\mu _B$ is the Bohr magneton, and $h$ is the Planck constant. 
The experimental signature is the change in the annihilation rates into 2$\gamma$ and 
3$\gamma$ final states caused by the Zeeman transition.

\section{Theoretical Resonance Line}
\label{sec:theoryofexperiment}
Our measurement directly determines $\Delta _{\mathrm{HFS}}$ 
using the theoretical resonance shape of $\Delta _{\mathrm{Zeeman}}$
 obtained using the density matrix of Ps spin states 
because the Breit-Rabi equation is not precise enough at ppm level. 
The following calculation is based on Refs.~\cite{MILLS-II,HUGHES-V}. 
The basis for the four spin eigenstates of Ps is defined as $\left( \psi _0, \psi _1, 
\psi _2, \psi _3 \right) \equiv \left( | S, S_z \rangle = | 0, 0 \rangle , 
| 1, 0\rangle , | 1, 1\rangle , | 1, -1 \rangle \right) $. We apply a magnetic field, 
\begin{equation}
{\bf B} (t) = B \hat{{\bf z}} + B _{\mathrm{RF}} \hat{{\bf x}} \cos \left( \omega t \right),
\end{equation}
where $\hat{{\bf z}}$ and $\hat{{\bf x}}$ are the unit vectors 
for the z and x directions respectively, 
$B_{\mathrm{RF}}$ is the magnetic field strength of the microwaves, 
$\omega$ is the frequency of the microwaves, and $t$ is the time since Ps is formed. 
The phase of the microwave is randomly distributed for each Ps in this experiment, 
but this effect on determination of $\Delta _{\mathrm{HFS}}$ is less than 0.1\,ppm 
so that arbitrary phase can be taken in the calculation.

The Hamiltonian $H$ including the Ps decay becomes
\begin{eqnarray}
\label{eq:hamiltonian}
H &=& h \Delta _{\mathrm{HFS}} (t) \times \nonumber \\
& & \left(
\begin{array}{cccc}
-\frac{1}{2} - \frac{i}{2} \gamma _{\mathrm{s}} & -q & r & -r \\
-q & \frac{1}{2}-\frac{i}{2} \gamma _{\mathrm{t}} & 0 & 0 \\
r & 0 & \frac{1}{2}-\frac{i}{2}\gamma _{\mathrm{t}} & 0 \\
-r & 0 & 0 & \frac{1}{2}-\frac{i}{2}\gamma _{\mathrm{t}} 
\end{array}
\right)  , 
\end{eqnarray}
where $r = g^{\prime } \mu _B B_{\mathrm{RF}} \cos \left( \omega t \right) / \left( \sqrt{2} h \Delta _{\mathrm{HFS}}(t) \right)$,
$\gamma _{\mathrm{s}} =  \Gamma ^{\prime} _{{\mathrm{p}}{\textrm{-}}{\mathrm{Ps}}} (t) /  \left( 2\pi \Delta _{\mathrm{HFS}}(t) \right) $, 
$\gamma _{\mathrm{t}} =  \Gamma ^{\prime} _{{\mathrm{o}}{\textrm{-}}{\mathrm{Ps}}} (t) / \left( 2\pi \Delta _{\mathrm{HFS}}(t) \right) $, 
$\Gamma ^{\prime} _{{\mathrm{p}}{\textrm{-}}{\mathrm{Ps}}} (t) = \Gamma _{{\mathrm{p}}{\textrm{-}}{\mathrm{Ps}}} + \Gamma _{\mathrm{pick}} (t)$, 
$\Gamma ^{\prime} _{{\mathrm{o}}{\textrm{-}}{\mathrm{Ps}}} (t) = \Gamma _{{\mathrm{o}}{\textrm{-}}{\mathrm{Ps}}} + \Gamma _{\mathrm{pick}} (t)$, 
and $\Gamma _{\mathrm{pick}} (t)$ is the 
pick-off ($ {\textrm{Ps}} + e^{-} \rightarrow 2 \gamma + e^{-}$) annihilation rate.
The time-dependence of $\Delta _{\mathrm{HFS}}$ and $\Gamma _{\mathrm{pick}}$ are caused by Ps thermalization, which is 
described later.
The $4\times 4$ density matrix $\rho(t)$ evolves with 
the time-dependent Schr\"{o}dinger equation, 
\begin{equation}
\label{eq:rho}
i \hbar \dot{\rho}  = H \rho - \rho H ^{\dagger }  ,
\end{equation}
where the $i$, $j$-element of $\rho(t)$ is defined as 
$\rho _{ij} (t) \equiv \langle \psi _{i} | 
\psi (t) \rangle \langle \psi (t) | \psi _{j} \rangle $ 
and the initial state is described as 
Eq.~(19) of Ref.~\cite{UNEUNE-1}. 
The 2$\gamma$ annihilation probability $(S_{2\gamma})$,  
and the 3$\gamma$ annihilation probability $(S_{3\gamma})$ 
are calculated 
between $t=t_0$ and $t=t_1$ as 
\begin{eqnarray}
\label{eq:s2gamma}
S_{2\gamma} &=& \int _{t_0} ^{t_1} \left( \Gamma ^{\prime} _{{\mathrm{p}}{\textrm{-}}{\mathrm{Ps}}}(t) \rho _{00}(t) + 
\Gamma _{\mathrm{pick}} (t) \sum _{i=1} ^{3} \rho _{ii} (t) \right) {\mathrm{d}}t , \\
S_{3\gamma } &=& \int _{t_0} ^{t_1}  \Gamma _{{\mathrm{o}}{\textrm{-}}{\mathrm{Ps}}}
\sum _{i=1} ^{3} \rho _{ii} (t)\, {\mathrm{d}}t .
\label{eq:s3gamma}
\end{eqnarray}
Furthermore, $S_{3\gamma }$ is divided into two components to calculate the experimental resonance line shape because 
of the different angular distribution of decay $\gamma$ rays from Ps between $|1,\pm 1 \rangle $ and $|1,0 \rangle $ states~\cite{3GAMMA_DECAY_ANGLE}.
The annihilation probability of $|1, \pm 1 \rangle$ state, 
$S_{|1,\pm 1\rangle } \equiv S_{|1, 1\rangle } + S_{|1, -1 \rangle }$, and the annihilation probability of 
$|1, 0 \rangle $ state, $S_{|1,0\rangle} $, are obtained by 
\begin{eqnarray}
\label{eq:s3gamma1}
S_{|1,\pm 1\rangle} &=& \int _{t_0} ^{t_1} \Gamma _{{\mathrm{o}}{\textrm{-}}{\mathrm{Ps}}} 
\left( \rho _{22} (t) + \rho _{33} (t) \right) {\mathrm{d}}t , \\
\label{eq:s3gamma0}
S_{|1, 0\rangle} &=& \int _{t_0} ^{t_1} \Gamma _{{\mathrm{o}}{\textrm{-}}{\mathrm{Ps}}}\, 
\rho _{11} (t)\, {\mathrm{d}}t .
\end{eqnarray}

\section{Ps Thermalization}
\label{sec:psthermalization}
Gas molecules are needed to form Ps in this experiment, but 
they make electric field around Ps which affects $\Delta _{\mathrm{HFS}}$.
This material effect (Stark effect) must be properly corrected to evaluate $\Delta _{\mathrm{HFS}}$ in vacuum. 
The Stark effect is estimated to be proportional to $n v (t) ^ {3/5}$, 
where $n$ is the number density of gas molecules
and $v (t) $ is the Ps mean velocity.
The $nv(t) ^{3/5}$ dependence of the Stark effect is calculated on the Lennard-Jones 
potential~\cite{ATOMIC_COLLISION}. 
The time dependence of $v (t)$ is caused by the Ps thermalization process. 
On the other hand, the measurement of the temperature dependence of the pick-off rate~\cite{PICK_TEMP} is consistent with an assumption that 
$\Gamma _{\mathrm{pick}} (t)$ is also proportional to $n v(t) ^{0.6}$ 
in {\it{i}}-C${\mathrm{_{4}H_{10}}}$ gas, which we use for Ps formation. 
It is obtained by fitting the data of Ref.~\cite{PICK_TEMP} by a power-law function of velocity, which 
results in an exponent of $\approx 0.6$. 
The uncertainty of the exponent is negligible for determination of $\Delta _{\mathrm{HFS}}$. 
The power-law dependence is indicated in Fig.~5 of Ref.~\cite{PICK_TEMP_THEO}.

According to the Ps thermalization model~\cite{NETSUKA-3}, $v (t)$ in gas is estimated as 
\begin{equation}
v(t) \approx \sqrt{ \frac{ 3 k T}{m _{\mathrm{Ps}} } } 
 \left( \frac{ 1 + A e ^{-bt} }{ 1 - A e^{-bt} } \right)  ,
\end{equation}
where $b = (16/3) \sqrt{2 / \pi } \sigma _{\mathrm{m}} n \sqrt{ m _{\mathrm{Ps}} k T} / M  $, 
$\sigma _{\mathrm{m}}$ is the momentum-transfer cross section of 
Ps collision with gas molecules, $m_{\mathrm {Ps}}$ is the Ps mass, $k$ is the 
Boltzmann constant, $T$ is the temperature of the gas, $M$ is the mass of the 
gas molecule, $A = ( \sqrt{E _{0}} - \sqrt{ (3/2) k T } ) / ( \sqrt{E _{0}} + \sqrt{ (3/2) k T})$, and 
$E _{0}$ is the initial kinetic energy of Ps. 
The thermalization parameters in {\it{i}}-C${\mathrm{_{4}H_{10}}}$ gas are measured 
to be $\left( E_0 = 3.1 ^{+1.0} _{-0.7} \,\textrm{eV}, \, \sigma_{\mathrm{m}} = 146 \pm 11 {\mathrm{\AA ^{2}}} \right)$ 
by DBS (Doppler Broadening Spectroscopy) technique~\cite{NETSUKA-5} in the range of 
0.15--1.52\,eV Ps kinetic energy.
However, the DBS result should not be applied for o-Ps whose kinetic energy is less than 0.17\,eV 
since $\sigma_{\mathrm{m}}$ depends on the kinetic energy of Ps.
As mentioned in Ref.~\cite{NETSUKA-5}, 
rovibrational excitations of the {\it{i}}-C${\mathrm{_{4}H_{10}}}$ molecule increase $\sigma _{\mathrm{m}}$ 
of Ps with kinetic energy above 0.17\,eV because {\it{i}}-C${\mathrm{_{4}H_{10}}}$ has a vibrational level at 0.17\,eV.
The `pick-off technique'~\cite{ASAI,JINNAI,KATAOKA}, which 
can access o-Ps with lower energy than 0.17\,eV, 
is a complementary method. 
This technique measures 
$\Gamma _{\mathrm{pick}} (t) / \Gamma _{\mathrm{o-Ps}} = 
(2\gamma \  {\mathrm{annihilation\ rate}}) / (3\gamma \ {\mathrm{annihilation\ rate}})$ as a function of time using $\gamma$-ray energy spectra. 
The thermalization can be measured by this method because 
$\Gamma _{\mathrm{pick}} (t)$ depends on the Ps velocity. 
The result of $\sigma _{\mathrm{m}} = 47.2 \pm 6.7\, {\mathrm{\AA ^{2}}}$ for o-Ps below 0.17\,eV 
has been obtained by our independent thermalization measurement using the `pick-off technique'.
In our analysis, the thermalization parameters from DBS measurement are used from $t = 0 $ to 
the time at which the kinetic energy of o-Ps reaches 0.17\,eV, and then the 
$\sigma _{\mathrm{m}}$ is changed to our value.

\section{Experimental Setup}
\label{sec:experimentalsetup}
\begin{figure}
\begin{center}
\includegraphics[width=0.45\textwidth]{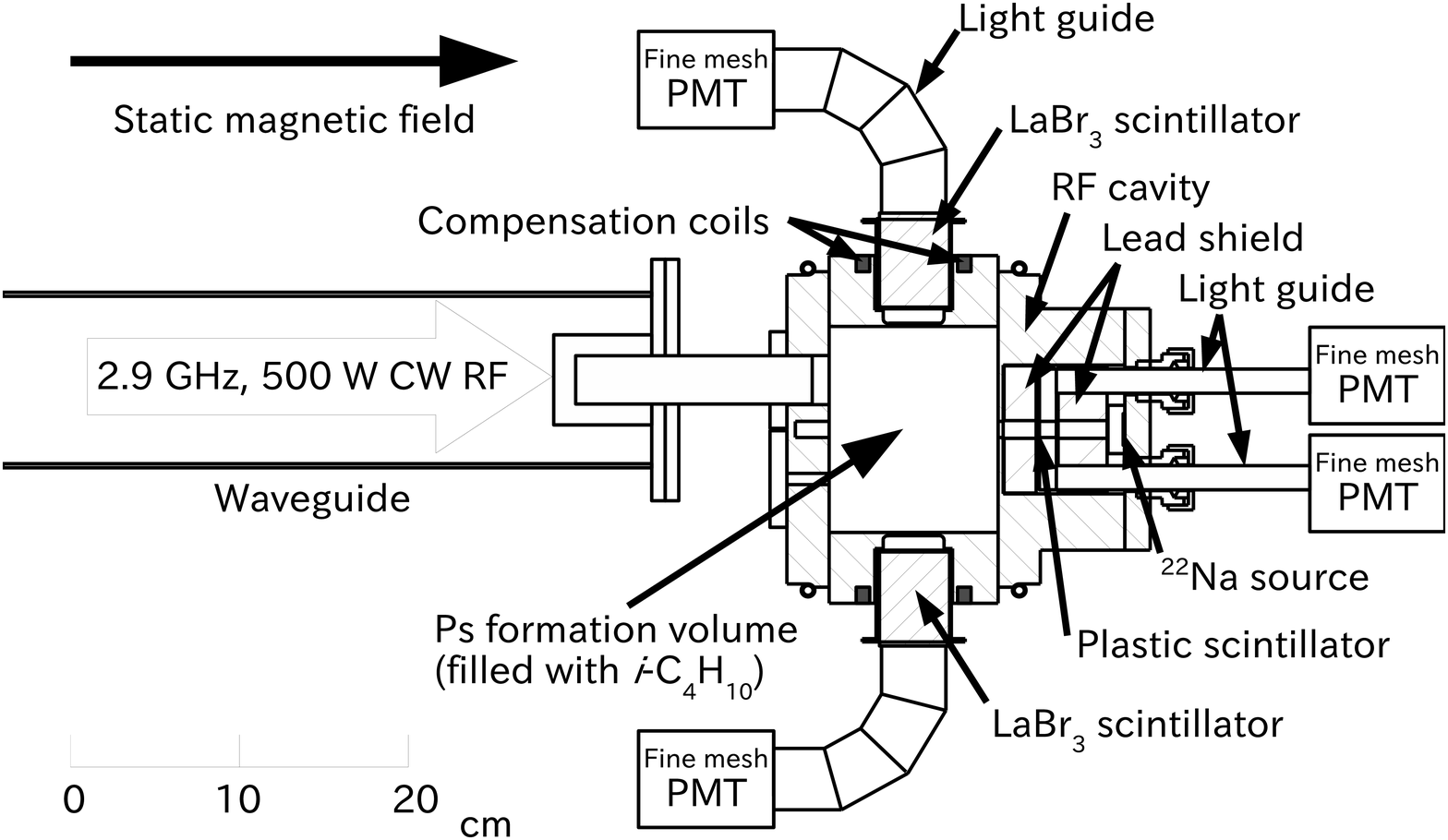}
\caption{\label{fig:schematic}Schematic diagram of the experimental setup (top view in magnet).}
\end{center}
\end{figure}

Figure \ref{fig:schematic} shows a schematic view of our experimental setup.
The timing information between Ps formation and decay is newly obtained in this experiment to 
investigate the non-thermalized Ps effect. 
The basic idea of the other setup is the same as the previous 
experiments~\cite{HUGHES-I,HUGHES-II,HUGHES-LETTER,MILLS-I,HUGHES-III,HUGHES-IV,MILLS-II,HUGHES-V}.

The positron source is 1\,MBq of $^{22}{\mathrm{Na}}$. 
A plastic scintillator 10\,mm in diameter and 0.1\,mm thick is used to tag positrons 
emitted from the source ($\beta$-tagging system). 
The scintillation light is detected by fine mesh photomultiplier tubes (PMTs) 
and provides a start signal corresponding to the 
time of Ps formation. The timing resolution is 1.2\,ns for 1 s.d. The positron 
enters the microwave cavity, forming Ps in pure ($> 99.9\%$) {\it{i}}-C${\mathrm{_{4}H_{10}}}$ gas contained therein.

Six $\gamma$ ray detectors are located around the microwave cavity to detect annihilation $\gamma$ rays. 
${\mathrm{LaBr_{3}(Ce)}}$ scintillators 
38.1\,mm in diameter and 50.8\,mm long are used, whose 
scintillation light is detected by fine mesh PMTs through UVT light guides as shown in Fig.~\ref{fig:schematic}. 
The energy resolution is 8\% FWHM at 511\,keV and the decay constant is as short as 16\,ns. 
The good energy resolution and fast response of LaBr$_3$(Ce) results in very good overall performance 
for measuring Zeeman transitions. 
In particular the acceptance of our setup is greatly increased by the good energy resolution, 
since 2$\gamma$ events are efficiently separated from 3$\gamma$ events with only energy information instead of a back-to-back geometry selection. 
The time spectrum between positron emission and $\gamma$-detection is measured
to improve the accuracy of the measurement of $\Delta _{\mathrm{HFS}}$.
The signal-to-noise ratio of the measurement is significantly improved by a factor of 20, since 
the prompt annihilation and p-Ps can be removed. 

A large bore superconducting magnet is used to produce a static magnetic field 
of $B \approx 0.866\,{\mathrm{T}}$. 
A bore diameter of the magnet is 800\,mm, and its length is 2\,m. 
The magnet is operated in persistent current mode, making the stability of the 
magnetic field better than $\pm 1\,{\mathrm{ppm}}$. With compensation coils surrounding 
the RF cavity, we achieve 1.5\,ppm RMS in uniformity of the magnetic field in the large volume of 
cylinder 40\,mm in diameter and 100\,mm long, where Ps are formed.
The magnetic field distribution is measured using a proton NMR magnetometer. 

Microwaves are produced by 
a local oscillator signal generator 
and amplified to 500\,W with a GaN amplifier. 
The input microwave power is monitored by power meters at two points, 
at an input waveguide and an antenna attached to the cavity.
The power is kept within 0.2\% short-term stability using a feedback system.
The microwave cavity is made with oxygen-free copper; 
inside of the cavity is a cylinder 128\,mm in diameter and 
100\,mm long. 
The $\gamma$ rays pass through the side wall of the cavity efficiently, since the 
thickness is only 1.5\,mm. 
The cavity is operated in the TM$_{110}$ mode. The resonant frequency is 2.856\,6\,GHz and 
the loaded quality factor $Q_L$ is 14,700.
The cavity is filled with pure {\it{i}}-C${\mathrm{_{4}H_{10}}}$ gas 
with a gas-handling system. At the first of every run, 
the cavity is pumped to the vacuum level of 
$10^{-4}\,{\mathrm{Pa}}$ and then the gas is filled to 0.129--1.366\,amagat\footnote{amagat is a unit of 
number density normalized by that of ideal gas at 0$^{\circ}$C, 1\,atm.}. 

\section{Analysis}
\label{sec:analysis}
Measurements were performed from July 2010 to March 2013.
In the overall period, the trigger rate was around 1.7\,kHz and the data acquisition rate was around 
910\,Hz. 
The signals from all PMTs were processed,
and the timing and the energy information were taken with NIM and CAMAC systems. 
The Zeeman transition was measured at various magnetic field strengths with fixed RF frequency and power.
The transition resonance lines were obtained at 11 gas density
(0.129, 0.133, 0.167, 0.232, 0.660, 0.881, 0.969, 1.193, 1.353, 1.358, and 1.366 amagat).
Data were taken at two different conditions, RF-ON and RF-OFF, at every gas density and magnetic field strength.
RF-ON data were taken with microwaves supplied. RF-OFF data were taken without microwave by switching off 
the signal generator and the amplifier. 

\begin{figure}
\begin{center}
\includegraphics[width=0.45\textwidth]{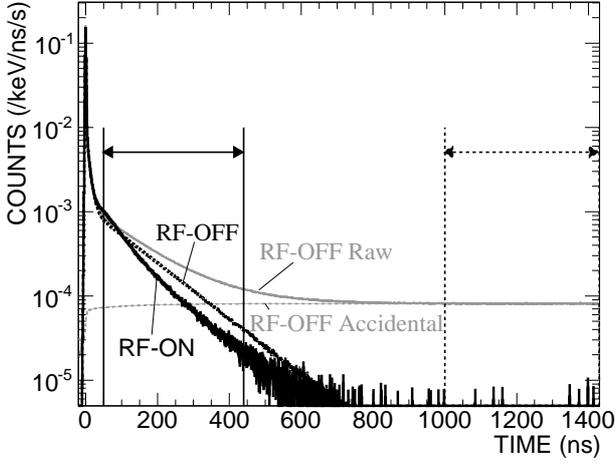}
\caption{\label{fig:timespectra}Timing spectra at 0.881\,amagat gas and 0.865\,733\,6\,T. 
The solid arrow shows the total timing window used for transition lines, 
and the dashed arrow shows the accidental timing window used for subtraction of energy spectra.
The accidental contribution has been already subtracted in the black `RF-OFF' and `RF-ON' lines.} 
\end{center}
\end{figure}

\begin{figure}
\begin{center}
\includegraphics[width=0.45\textwidth]{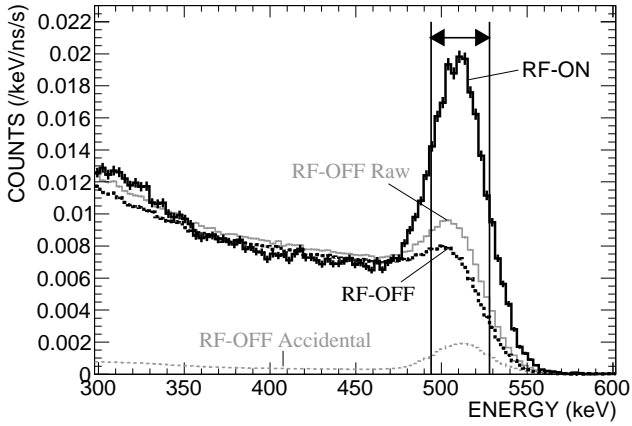}
\caption{\label{fig:energyspectra}Energy spectra
at 0.881\,amagat gas  and 0.865\,733\,6\,T in the timing window of 50--60\,ns. 
The accidental contribution has been already subtracted in the black `RF-OFF' and `RF-ON' lines. 
The transition lines are obtained by comparing the areas 
of RF-ON and RF-OFF inside the energy window indicated by the arrow.}
\end{center}
\end{figure}

\begin{figure}
\begin{center}
\begin{minipage}{0.5\textwidth}
\begin{center}
	\includegraphics[width=0.9\textwidth]{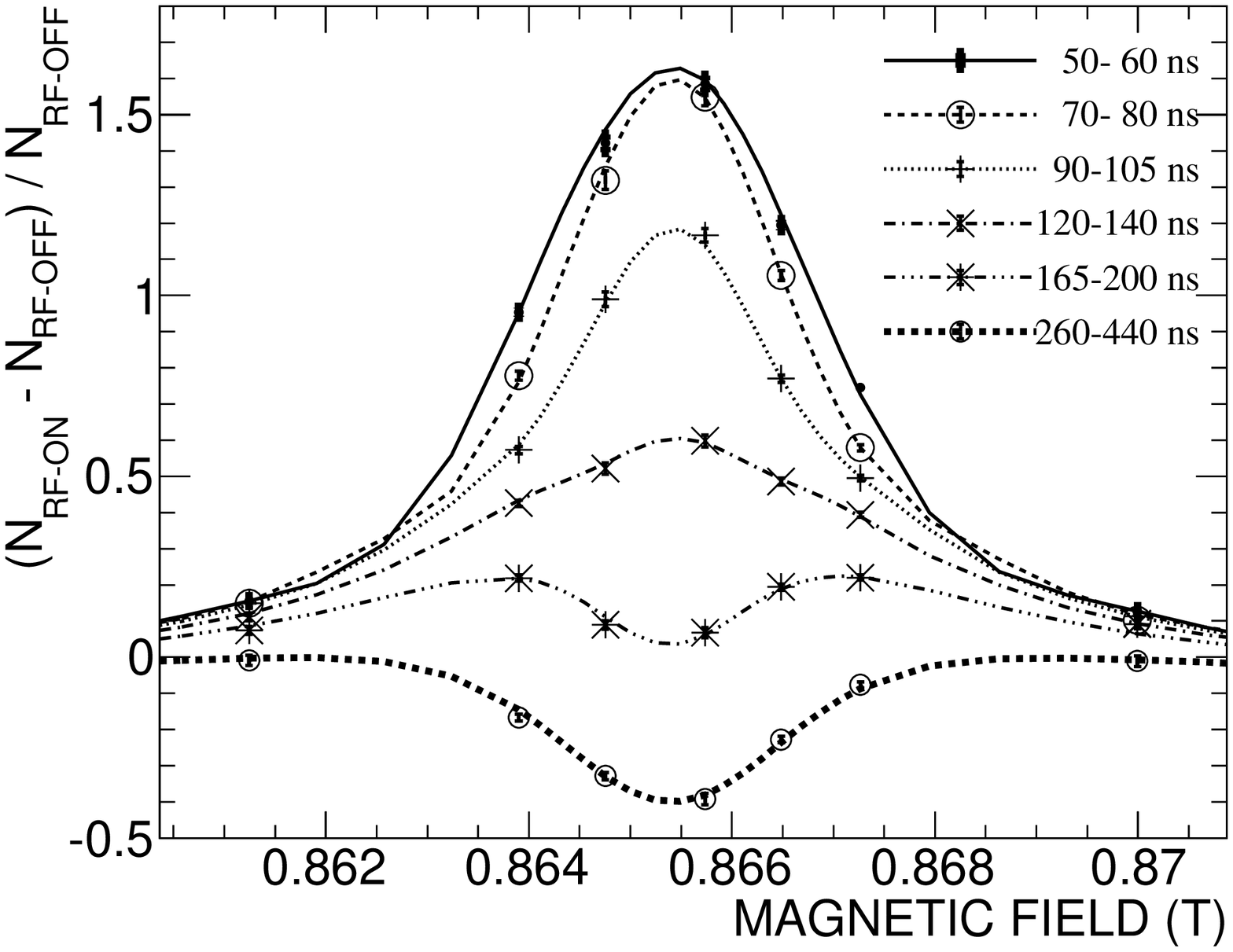}
	\label{fig:transition_01}
\end{center}
\end{minipage}
\begin{minipage}{0.5\textwidth}
\begin{center}
	\includegraphics[width=0.9\textwidth]{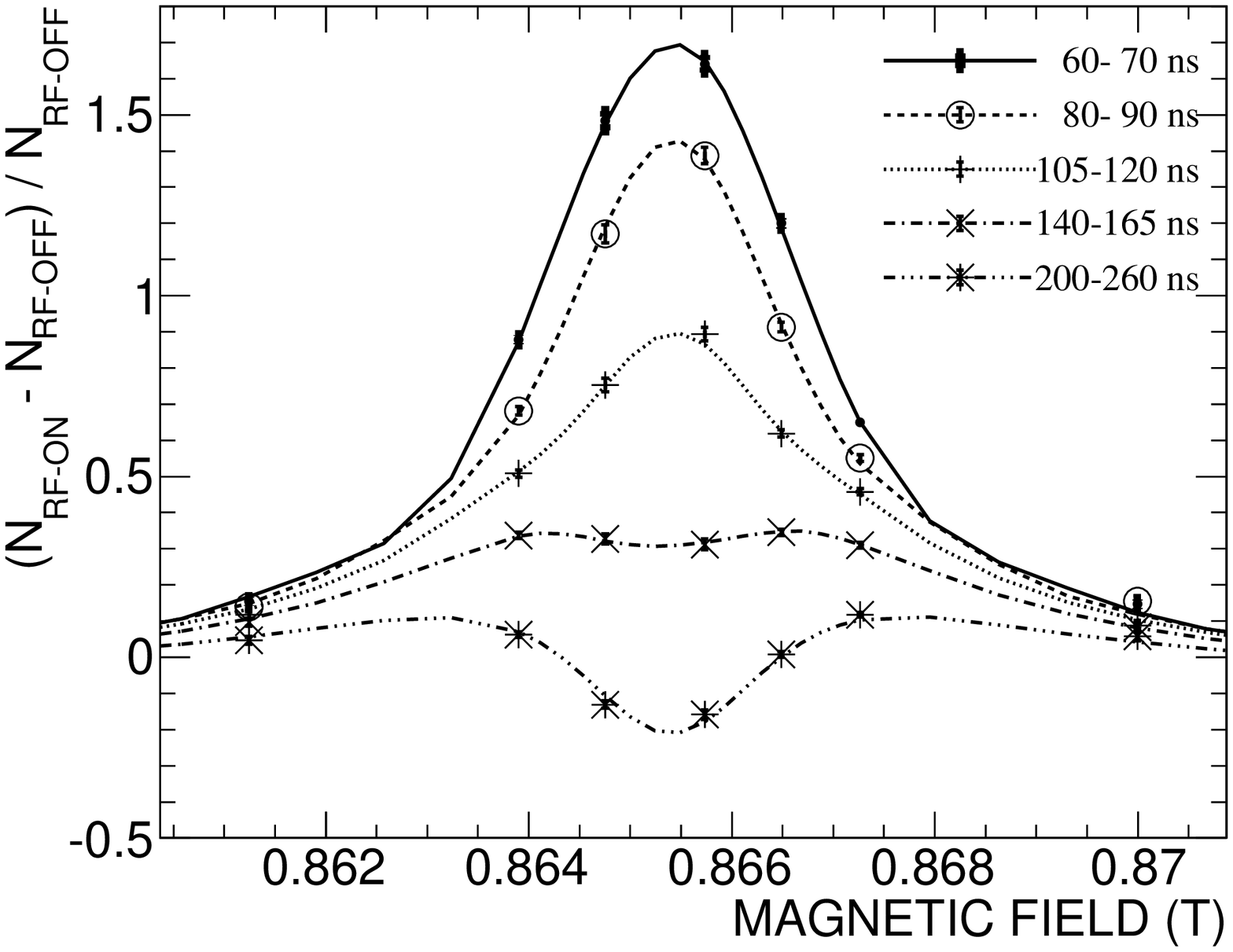}
	\label{fig:transition_02}
\end{center}
\end{minipage}
\caption{\label{fig:resonance_line}Resonance lines at 0.881\,amagat gas. 
The markers with error bars indicate obtained data, and the lines indicate the best-fit result. 
Eleven lines are divided into two figures for improvement of visibility.}
\end{center}
\end{figure}

Figure~\ref{fig:timespectra} shows a typical timing spectra between the $\beta$-tag 
and the $\gamma$-signal.
The timing spectra without accidental contribution are obtained by 
subtracting the accidental spectra from the raw timing spectra.
The accidental spectra are calculated using Ref.~\cite{TIME_COLEMAN}. 
The difference from a simple exponential shape is because of a change of an efficiency of accidental events, which 
depends on the true signal shape.
The true timing spectrum with true rate is obtained by correcting the suppression 
caused by dead time of electronics.
The difference of the slope between RF-ON and RF-OFF is caused by the Zeeman transition.
A timing window of 50--440\,ns is applied to select o-Ps events. 
The window is divided into 11 sub-windows in our analysis, and 
the time evolution of the Zeeman transition is confirmed.
The energy spectra are obtained by 
subtracting the accidental contribution from the raw spectra as shown in Fig.~\ref{fig:energyspectra}.
The accidental energy spectra are estimated using the energy spectra in the timing window of $t = $1,000--1,430\,ns. 
The resonance lines are obtained by 
$\left( N_{{\mathrm{RF}}{\textrm{-}}{\mathrm{ON}}} - N_{{\mathrm{RF}}{\textrm{-}}{\mathrm{OFF}}} \right) / N_{{\mathrm{RF}}{\textrm{-}}{\mathrm{OFF}}} $ 
as a function of the static magnetic field strength,
where $N_{{\mathrm{RF}}{\textrm{-}}{\mathrm{ON}}}$ is the counting rate of the events in the energy window of 
$511\,{\textrm{keV}} \pm 1 \,\textrm{s.d.} (\approx 17\,{\mathrm{keV}})$ of RF-ON, and 
$N_{{\mathrm{RF}}{\textrm{-}}{\mathrm{OFF}}}$ is that of RF-OFF.
Typical resonance lines obtained are shown in Fig.~\ref{fig:resonance_line}. 

Resonance lines are fitted by the following function $F(t,n,B)$:
\begin{eqnarray}
F(t,n,B) &=& D_{1}(n) 
\frac{R_{{\mathrm{RF}}{\textrm{-}}{\mathrm{ON}}}(t,n,B) - R_{{\mathrm{RF}}{\textrm{-}}{\mathrm{OFF}}}(t,n,B)}
{R_{{\mathrm{RF}}{\textrm{-}}{\mathrm{OFF}}}(t,n,B)} \nonumber \\
 &+& D_{2}(n) , 
\label{eq:fitfunc}\\
R(t,n,B) &\equiv& \epsilon (n) S_{2\gamma} (t,n,B) + S_{|1,\pm 1\rangle} (t,n,B) \nonumber \\
 &+& \epsilon ^{\prime} (n) S_{|1,0 \rangle} (t,n,B) \label{eq:fitfuncR} ,
\end{eqnarray}
where $n$ is the number density of gas molecules, $D_{1} (n)$ is a normalization factor, 
$D_{2} (n)$ is an offset, $\epsilon (n)$ and $\epsilon^{\prime} (n)$ are the ratios of detection efficiencies 
of ${2\gamma}$ and $|1,0 \rangle$ decay, respectively, normalized by that of the $| 1,\pm 1\rangle$ decay.
$S_{2\gamma}$, $S_{|1,\pm 1 \rangle}$, and $S_{|1,0 \rangle}$ are calculated numerically 
from Eqs. (\ref{eq:s2gamma}), (\ref{eq:s3gamma1}), and (\ref{eq:s3gamma0}), respectively. 
In the fitting process, 
$D _{1} (n)$ and $D _{2} (n)$ are treated as free parameters for each gas density 
because of the following three reasons.
The first one is normalization of the counting rate of RF-ON and that of RF-OFF. It is caused by 
the fact that {\it{i}}-C${\mathrm{_{4}H_{10}}}$ slightly absorbs microwaves which makes the gas temperature high (the density low).
The second one is the contribution from Ps formed in the region where microwaves are not supplied. 
The third one is the difference of the second one between RF-ON and RF-OFF.
$\epsilon (n)$ and $B_{\mathrm{RF}} (n)$ are also treated as free parameters since the
distribution of Ps formation position in the cavity depends on the gas density and this 
dependency makes the detection efficiency and the effective $B_{\mathrm{RF}}$ depend on the gas density.
A typical value of $\epsilon$ is 6.5.
The effective $B_{\mathrm{RF}}$ is typically decreased by about $10\%$ 
from maximum value (typically 15\,G) because of the distribution.
$\epsilon ^{\prime}$ is estimated by GEANT4~\cite{GEANT4-1, GEANT4-2} 
Monte Carlo simulation in which all the materials are 
reproduced and Ps formation position is also simulated. 
A typical value of $\epsilon ^{\prime}$ is 1.139.
The uncertainty from this MC estimation is negligible because the contribution of 
the $|1,0 \rangle$ state is small. 
The polarization of positron which forms Ps is also estimated by GEANT4 MC simulation. 
Estimated values fall within the range of 0.23 at low gas density and 0.42 at high gas density.
Comparisons with unpolarized and completely polarized estimations have been performed, 
but the shifts of the final fitted Ps-HFS value 
has been less than 0.2\,ppm. It shows that the uncertainty of this MC estimation is also negligible. 
The Doppler broadening effect is taken into account by a convolution with 
the Gaussian distribution of $\omega$ with a s.d. of $\omega \sqrt{kT/(m_{\mathrm{Ps}} c^2)}$, 
where $c$ is the speed of light in vacuum. 

The time dependence of $\Delta _{\mathrm{HFS}} (t)$ and $\Gamma _{\mathrm{pick}} (t)$ are estimated 
using the following thermalization effect and they are taken into account 
in the evolution of $S_{2\gamma}$, $S_{|1,\pm 1\rangle}$, and $S_{|1,0 \rangle}$ as
\begin{eqnarray}
\Delta _{\mathrm{HFS}} (n,t) &=& \Delta _{\mathrm{HFS}} ^{0} - C n v (t) ^{3/5} , 
\label{eq:fittingC}\\
\Gamma _{\mathrm{pick}} (n,t) &=& \Gamma_{\mathrm{pick}} (n, \infty) \times 
\left( \frac{ v(t) }{ v(\infty) } \right) ^{0.6}  ,
\end{eqnarray}
where $\Delta _{\mathrm{HFS}} ^{0}$ is Ps-HFS in vacuum and $C$ is a constant. 
$\Delta _{\mathrm{HFS}} ^{0}$ and $C$ are common free parameters of fitting for 
all data points. 
$\Gamma _{\mathrm{pick}} (n, \infty)$ is determined by fitting the RF-OFF timing spectra 
for each gas density with the following equation $N(t)$ including Ps thermalization effect~\cite{ASAI,JINNAI}:
\begin{eqnarray}
N(t) &=& N_{0} \exp \left[ - \Gamma_{{\mathrm{o}}{\textrm{-}}{\mathrm{Ps}}} 
\int _{0} ^{t} \left( 1 + \frac{ \Gamma _{\mathrm{pick}} (t^{\prime}) }
{\Gamma _{{\mathrm{o}}{\textrm{-}}{\mathrm{Ps}}}} \right) \mathrm{d}t^{\prime} \right] \nonumber \\
&+& N_{1} \exp \left[ - \Gamma_{|+ \rangle} \int _{0} ^{t} 
\left( 1 + \frac{ \Gamma _{\mathrm{pick}} (t^{\prime}) } {\Gamma _{|+ \rangle}} \right) \mathrm{d}t^{\prime} \right] ,
\end{eqnarray}
where $N_{0}$ and $N_{1}$ are normalization constants, 
$\Gamma_{|+\rangle}$ is the decay rate of Ps for the highest energy state with the longer lifetime of 
the two mixed states of $|1,0 \rangle$ and $|0,0 \rangle$. 
Another component of the mixed states is ignored because of its short lifetime. 

As shown in Fig.~\ref{fig:resonance_line}, the data points are well fitted by 
Eq.~(\ref{eq:fitfunc}). 
By fitting all of our data points (11 gas density $\times$ 11 timing windows $\times$ 4--7 
magnetic field strengths) simultaneously, the best-fit value of 
\begin{equation}
\Delta _{\mathrm{HFS}} ^{0} = 203.394\,2(16) \, {\textrm{GHz}} 
\label{eq:hfsstat}
\end{equation}
is obtained with $\chi ^{2} / {\textrm{ndf}} = 633.3/592$ and a $p$-value of 0.12. 
Time evolution of some parameters: $\rho _{00} (t)$, $v(t)/c$, 
$\Gamma _{\mathrm{pick}} (t)$, and $\Delta _{\mathrm{HFS}} (t)$, at various gas density are shown 
in Figs.~\ref{fig:rho00}--\ref{fig:hfs}.

In order to evaluate the non-thermalized Ps effect, which was not considered in the previous 
experiments, fitting without taking into account the time evolution of
$\Delta _{\mathrm{HFS}}$ and $\Gamma _{\mathrm{pick}}$ is performed.
The fitted Ps-HFS value with an assumption that Ps is well thermalized results in 
$203.392\,2(16) \, {\textrm{GHz}}$. 
Comparing it with Eq. (\ref{eq:hfsstat}), the non-thermalized o-Ps effect 
is evaluated to be as large as $10 \pm 2\,{\textrm{ppm}}$ in the timing window we used. 
This effect might be larger if no timing window is applied, 
since it depends on the timing window used for the analysis. 
In the timing window of 0--50\,ns, which we do not use for the analysis, Ps-HFS is dramatically changing 
because Ps is not well thermalized and Ps velocity is still rapidly changing.

\begin{table}
\begin{center}
\caption{\label{tab:systematicerror}Summary of systematic errors.}
\begin{tabular}{lr}
\hline \hline
Source & Errors in $\Delta _{\mathrm{HFS}}$ (ppm) \\
\hline
{\it Material Effect:} & \\
~~~{\small{o-Ps pick-off}} & 3.5 \\
~~~{\small{Gas density measurement}} & 1.0 \\
~~~{\small{Temperature  measurement}} & 0.1 \\
~~~{\small{Spatial distribution of density}} & \\
~~~{\small{and temperature in the RF cavity}} & 2.5 \\
~~~{\small{Thermalization of Ps;}} & \\
~~~{\small{~~~Initial kinetic energy $E_0$}} & 0.2 \\
~~~{\small{~~~DBS result $\sigma _{\mathrm{m}}$}} & 0.5 \\
~~~{\small{~~~pick-off result $\sigma _{\mathrm{m}}$}} & 1.8 \\
{\it Magnetic Field:} & \\
~~~{\small{Non-uniformity}} & 3.0 \\
~~~{\small{Offset and reproducibility}} & 1.0 \\
~~~{\small{NMR measurement}} & 1.0 \\
{\it RF System:} & \\
~~~{\small{RF power}} & 1.2 \\
~~~{\small{$Q_L$ value of RF cavity}} & 1.2 \\
~~~{\small{RF frequency}} & 1.0 \\
~~~{\small{Power distribution in the cavity}} & $< 0.1$ \\
{\it Others:} & \\
~~~{\small{Choice of timing window}} & 1.8 \\
~~~{\small{Choice of energy window}} & 0.6 \\
~~~{\small{Polarization of $e^{+}$}} & $< 0.2$ \\
~~~{\small{Phase of microwaves}} & $< 0.1$ \\
~~~{\small{o-Ps lifetime}} & $< 0.1$ \\
~~~{\small{p-Ps lifetime}} & $< 0.1$ \\
\hline
Quadrature sum & 6.4 \\
\hline \hline
\end{tabular}
\end{center}
\end{table}

Systematic errors are summarized in Table~\ref{tab:systematicerror}.
The largest contribution is an uncertainty of the material effect. 
An uncertainty of o-Ps pick-off rate ($\Gamma _{\mathrm{pick}} ( n, \infty )$) 
is estimated by taking the error of the fitting of the o-Ps decay curve. 
The uncertainty of the gas density is computed from the uncertainties of the gas pressure and 
temperature as 0.2\%, resulting in 1.0\,ppm uncertainty in $\Delta _{\mathrm{HFS}}$. 
The uncertainty of the gas temperature is estimated to be 0.1\,K, which corresponds to 0.1\,ppm in $\Delta _{\mathrm{HFS}}$.
In order to estimate a systematic uncertainty from the spatial distributions of gas density and temperature in the RF cavity,
these distributions with an extreme condition of no gas convection are estimated.
It is assumed that
the strength of RF power absorbed by the gas is proportional to the energy density of electric field of TM$_{110}$ mode. 
As a result, 
the gas temperature distribution of $\approx 170\,{\mathrm{K}}$ range 
is produced in the RF cavity, and the fitting result of 
$\Delta _{\mathrm{HFS}}$ shifts by $+2.5\,\mathrm{ppm}$. 
This shift is conservatively considered as a systematic error. 
The uncertainty of Ps thermalization effect is estimated by the errors of the thermalization parameters.

The second largest contribution is an uncertainty of the static magnetic field.
Distribution of the static magnetic field is measured by the NMR magnetometer with the same setup as 
Ps-HFS measurement for twice (before and after the measurement). 
The results of the two measurements are consistent with each other and the 
non-uniformity is weighted by the RF magnetic field strength and distribution of 
Ps formation position, which results in 1.5\,ppm RMS inhomogeneity. 
The strength of the static magnetic field is measured outside of the RF cavity during the run.
An offset value at this point is measured during the measurement of the magnetic field 
distribution, and its uncertainty including reproducibility is 0.5\,ppm.
The precision of magnetic field measurement is 0.5\,ppm, which comes from 
the polarity-dependence of the NMR probe. These uncertainties are doubled because 
$\Delta_{\mathrm{HFS}}$ is approximately proportional to the square of the static magnetic field strength.

Uncertainties related to RF system are estimated by 
uncertainties of all the RF parameters included in the fitting; 
power, $Q_L$ value of the cavity, and frequency. 
A long-term stability of 0.06\% and a relative uncertainty of measurement of 0.08\% are 
concerned about the power, which results in 0.10\% total uncertainty, corresponding to 1.2\,ppm 
error in $\Delta _{\mathrm{HFS}}$. 
A long-term stability of 0.08\% and a relative uncertainty of measurement of 0.06\% are 
concerned about the $Q_L$ value, which results also in 0.10\% total uncertainty, 
corresponding to 1.2\,ppm error in $\Delta _{\mathrm{HFS}}$. 
A long-term stability of 0.8\,ppm and an absolute uncertainty of 0.6\,ppm are 
concerned about the frequency, which results in 1.0\,ppm total uncertainty, 
corresponding to 1.0\,ppm error in $\Delta _{\mathrm{HFS}}$. 
In our final global-fitting, Eq.~(\ref{eq:rho}) is solved for a given average RF power. 
However, depending on their position the Ps see different power and the final results should be 
given by an average of several Eq.~(\ref{eq:rho}) for different power. 
A fitting with this method has been performed to estimate this effect. 
The distribution of Ps formation position is estimated using GEANT4 MC simulation. 
Time dependence of RF power for each Ps is ignored because the diffusion length of Ps within its lifetime 
is less than 1\,mm and there is no large difference of RF power at this distance. 
A free parameter of proportionality coefficient to this distribution has been used instead of 
$B_{\mathrm{RF}}$. 
The shift of $\Delta _{\mathrm{HFS}}$ has been less than 0.1\,ppm. 
An estimation including spatial distribution of gas density has also been performed and 
the shift has also been less than 0.1\,ppm. 
This shift is considered as a systematic error. 

Other systematic uncertainties are related to the analysis. 
Fittings with the starting time of 40\,ns and 60\,ns with the fixed fitting end time of 440\,ns are performed 
in order to study a systematic error of the choice of the timing window. 
Fittings with the ending time of 260\,ns and 620\,ns with the fixed fitting start time of 50\,ns are also performed. 
The maximum shift in $\Delta _{\mathrm{HFS}}$ is 1.8\,ppm and it is considered as a systematic error. 
The gain and offset of the detectors 
are calibrated every 10 minutes and their uncertainties are negligible.
Analysis with energy window of 
$511\,{\mathrm{keV}} \pm 1.5\,{\mathrm{s.d.}} ( \approx 26\,{\mathrm{keV}}) $ has 
been performed, and the result has shifted by 0.6\,ppm. 
This shift is taken into account as a systematic error of the choice of the energy window. 
Other systematic errors from detectors are considered to be cancelled out 
by the subtraction of RF-OFF data from RF-ON data and the normalization by RF-OFF data. 
The uncertainties of lifetime measurements of Ps affect $\Delta _{\mathrm{HFS}}$ by 
less than 0.1\,ppm. 
An effect of excited states can be estimated using the Hamiltonian as shown in Ref.~\cite{EXCITED} 
and it is negligible. 
Other systematic errors such as 
the motional Zeeman and Stark effects, the spin-conversion quenching 
of Ps, the quadratic Zeeman effect, and smaller correction to $g$ factor are negligible. 

\begin{figure}
\begin{center}
\includegraphics[width=0.45\textwidth]{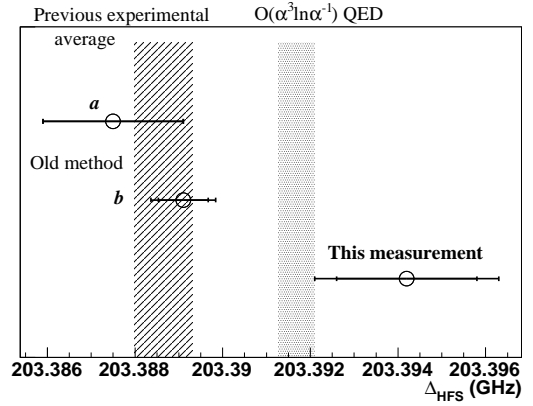}
\caption{\label{fig:history}Summary of $\Delta _{\mathrm{HFS}}$ measurements from past experiments and this work. 
The circles with error bars are the experimental data ($a-$\cite{MILLS-II}, $b-$\cite{HUGHES-V}), 
the hatched band is the average of the previous experiments ($a$ and $b$), 
and the dotted band is the QED calculation~\cite{HFS-ORDER3-KNIEHL,HFS-ORDER3-MELNIKOV,HFS-ORDER3-HILL}.}
\end{center}
\end{figure}

The systematic errors discussed above are regarded as independent, 
and the total systematic error is calculated to be their 
quadrature sum. 
When the non-thermalized Ps effect is included, our final result with the systematic errors is 
\begin{equation}
\Delta_{\mathrm{HFS}} = 203.394\,2 \pm 0.001\,6 ({\mathrm{stat.}}) 
\pm 0.001\,3 ({\mathrm{sys.}}) \, {\mathrm{GHz}}.
\end{equation}
A summary plot of $\Delta_{\mathrm{HFS}}$ measurements is shown in Fig.~\ref{fig:history}.
Our result favors the QED calculation within 1.2 s.d., although it disfavors the 
previous experimental average by 2.6 s.d.

\section{Conclusion}
\label{sec:conclusion}
A new precision measurement of Ps-HFS free from possible common uncertainties 
from Ps thermalization effect was performed to check the Ps-HFS discrepancy.
The effect of non-thermalized o-Ps was evaluated to be as large as $10 \pm 2\,{\mathrm{ppm}}$ in a timing window we used. 
This effect might be larger than 10\,ppm if no timing window is applied, since it depends on timing window.
Including this effect, our new experimental value results in 
$\Delta_{\mathrm{HFS}} = 203.394\,2 \pm 0.001\,6 ({\mathrm{stat.}},8.0\,{\mathrm{ppm}}) 
\pm 0.001\,3 ({\mathrm{sys.}},6.4\,{\mathrm{ppm}}) \, {\mathrm{GHz}}$. 
It favors the $O(\alpha ^{3} \ln \alpha ^{-1}) $ QED calculation within 1.2 s.d., 
although it disfavors the previous measurements by 2.6 s.d.

Sincere gratitude is expressed to Dr.~T.~Suehara (Kyushu~U.), Mr.~Y.~Sasaki, 
Mr.~G.~Akimoto (U.~Tokyo), Prof.~A.~P.~Mills,~Jr. (UC Riverside), Dr.~H.~A.~Torii and Dr.~T.~Tanabe (U.~Tokyo) for useful discussions. 
We warmly thank facilities and the entire members of 
the Cryogenics Science Center at KEK 
without whose excellent support this experiment could not have been 
successfully performed. 
This work was supported by JSPS KAKENHI Grant Number 23340059.

\appendix
\section{Time Evolution of Parameters at various gas density}
\label{sec:appendix1}
\begin{figure}
\begin{center}
\includegraphics[width=0.45\textwidth]{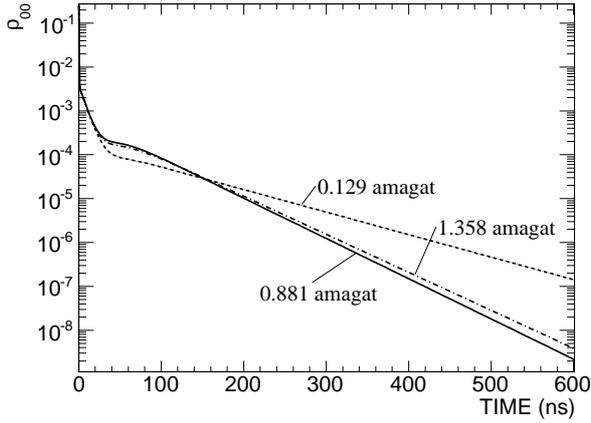}
\caption{\label{fig:rho00}Time evolution of $\rho _{00}$.}
\end{center}
\end{figure}

\begin{figure}
\begin{center}
\includegraphics[width=0.45\textwidth]{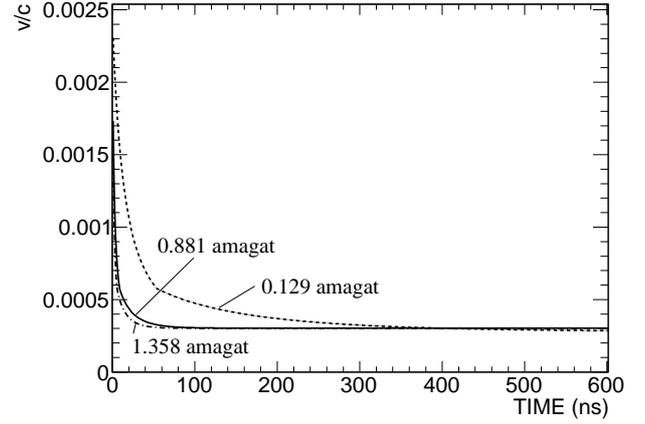}
\caption{\label{fig:velocity}Time evolution of Ps velocity normalized by $c$.}
\end{center}
\end{figure}

\begin{figure}
\begin{center}
\includegraphics[width=0.45\textwidth]{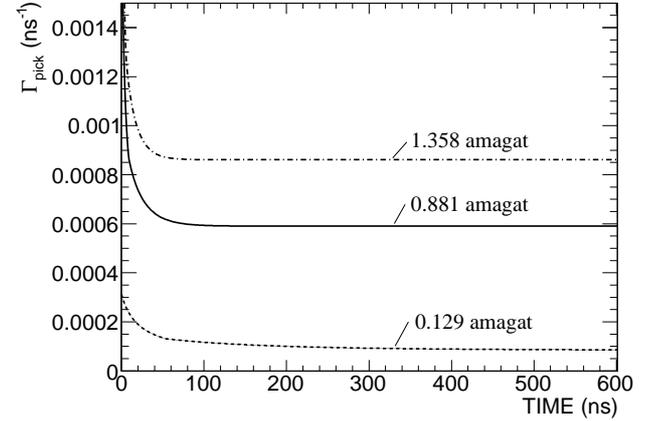}
\caption{\label{fig:pickoff}Time evolution of pick-off annihilation rate.}
\end{center}
\end{figure}

\begin{figure}
\begin{center}
\includegraphics[width=0.45\textwidth]{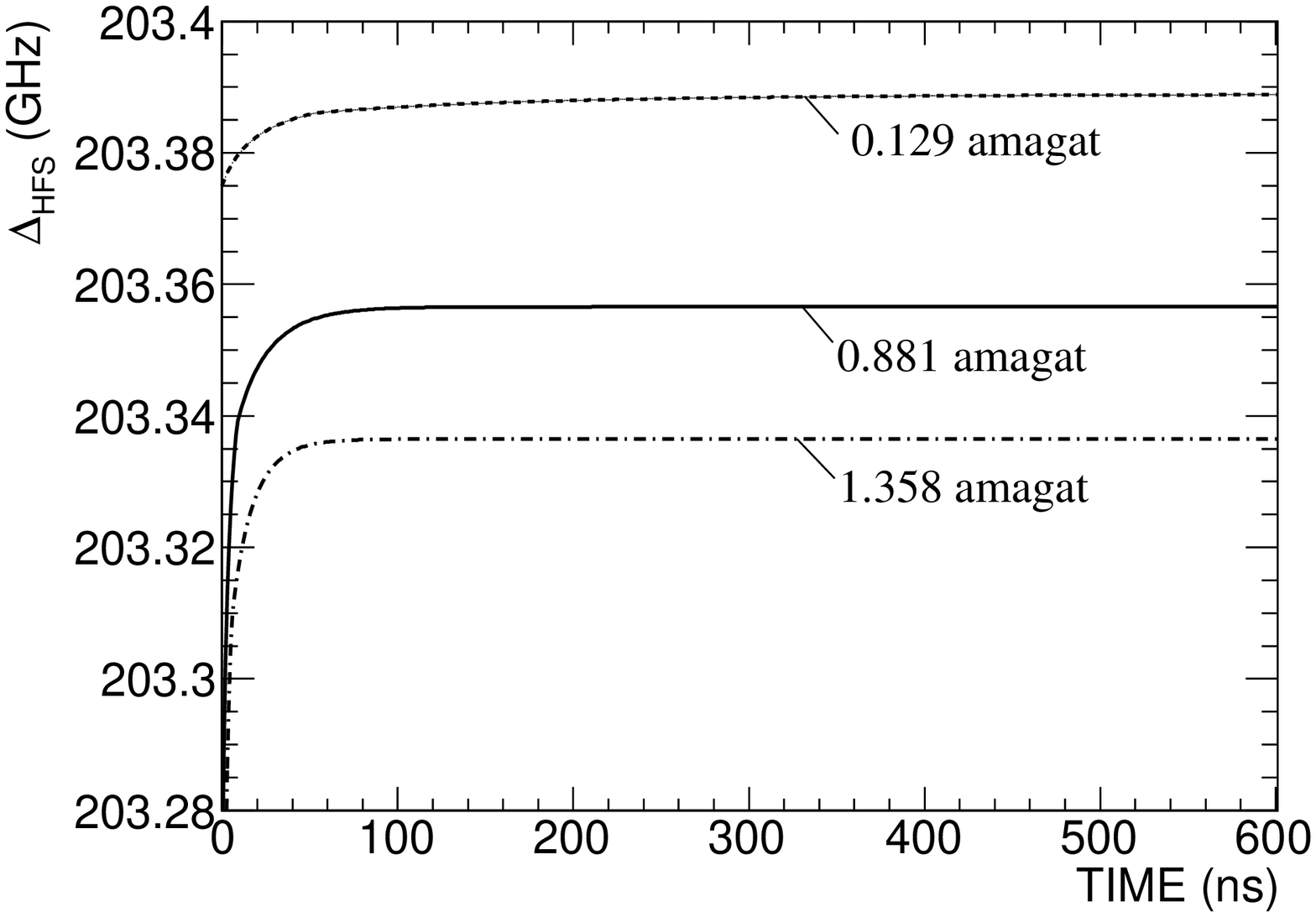}
\caption{\label{fig:hfs}Time evolution of $\Delta _{\mathrm{HFS}}$.}
\end{center}
\end{figure}

Time evolution of some parameters: $\rho _{00} (t)$, $v(t)/c$, 
$\Gamma _{\mathrm{pick}} (t)$, and $\Delta _{\mathrm{HFS}} (t)$, at several gas density  
using our final fitting results are shown in Figs.~\ref{fig:rho00}--\ref{fig:hfs}.
The graphs are drawn at the static magnetic field strengths of the nearest data points to the 
centers of the resonances at 0.129, 0.881, and 1.358\,amagat gas density.
Fig.~\ref{fig:rho00} shows the time evolution of one component $\rho _{00} (t)$ of $4 \times 4$ 
density matrix $\rho$ of Ps spin states. The 2$\gamma$ annihilation rate is mainly proportional 
to this function. The graphs are drawn with RF-ON condition, and the shape depends on the microwave power. 
At low gas density, the measurements were performed with low microwave power to avoid discharge. 
Fig.~\ref{fig:velocity} shows the time evolution of the normalized Ps velocity $v(t)/c$. 
It shows that the thermalization takes much time at low gas density. Kinks where 
$\sigma _{\mathrm{m}}$ changes because Ps energy across the 0.17\,eV threshold are shown. 
Fig.~\ref{fig:pickoff} shows the time evolution of the pick-off annihilation rate $\Gamma _{\mathrm{pick}} (t)$. 
It shows the $n v(t) ^{0.6}$ dependence and the Ps thermalization is clearly seen. 
The `pick-off technique' originally measures this function to obtain the thermalization parameters. 
Fig.~\ref{fig:hfs} shows the time evolution of Ps-HFS $\Delta _{\mathrm{HFS}} (t)$.  
A dramatic change of $O(100)\,{\mathrm{ppm}}$ is shown in the timing range earlier than 50\,ns which 
we do not use for the analysis. 
A slow change of $O(10)\,{\mathrm{ppm}}$ is also shown at low gas density. 
These are the effect of non-thermalized o-Ps on $\Delta _{\mathrm{HFS}}$. 

\section{$\Delta _{\mathrm{HFS}}$ versus gas density}
\label{sec:appendix2}
\begin{figure}
\begin{center}
\includegraphics[width=0.45\textwidth]{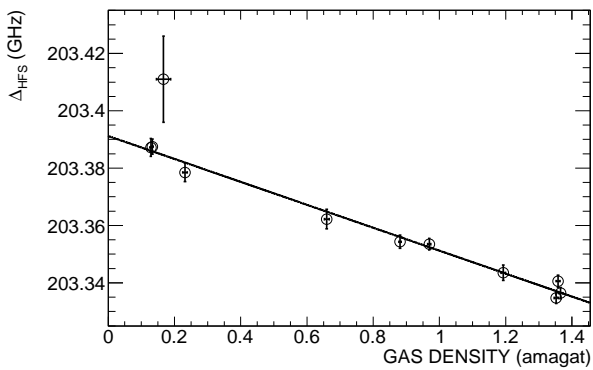}
\caption{\label{fig:old}$\Delta _{\mathrm{HFS}}$ at each gas density. The circles with error bars are 
the data, and the solid line is the best-fit with a linear function.}
\end{center}
\end{figure}

Completely separate analysis which determine $\Delta _{\mathrm{HFS}}$ value at each gas density 
has been performed to provide additional insight into the complete experimental data set and 
confirm their quality. 
Figure~\ref{fig:old} shows the result. 
It is obtained by fitting the resonance lines at each gas density without considering the 
time evolution of $\Delta _{\mathrm{HFS}}$, {\it{i.e.}} 
$\Delta _{\mathrm{HFS}}$ is treated as a constant at each gas density 
instead of using Eq.~(\ref{eq:fittingC}).  
This method is similar to the method used in the previous experiments except that 
our data use timing information, which was not taken in the previous measurement, 
and 11 resonance lines within 50--440\,ns timing window are simultaneously fitted at each gas density. 
It is impossible to include the time evolution of 
$\Delta _{\mathrm{HFS}}$ in this method. 
It is evident that the data fluctuations from the linear-fit function 
are reasonable compared to the error bars. It is important to say that 
determination of $\Delta _{\mathrm{HFS}}$ using our data needs our new global-fitting method 
to treat the time evolution correctly. 

\bibliographystyle{model1-num-names}
\bibliography{ishida}







\end{document}